\documentclass[twocolumn,showpacs,superscriptaddress,amsmath,amssymb,prl]{revtex4-1}
\usepackage[colorlinks,linkcolor=blue,anchorcolor=blue,citecolor=blue,urlcolor=blue]{hyperref}
\usepackage{graphicx}
\usepackage{epsfig}
\usepackage{dcolumn}
\usepackage{bm}
\usepackage{overpic}
\usepackage{color}
\usepackage{multirow}
\usepackage{subfigure}
\usepackage{lineno}
\usepackage{verbatim}
\usepackage{cleveref}
\usepackage{amsthm,amsmath,amssymb}
\usepackage{mathrsfs}
\usepackage{booktabs}
\newcommand\jpsi{J/\psi}
\newcommand\psip{\psi(3686)}
\newcommand\etap{\eta'}

\newcommand\kpkm{K^+K^-}

\begin{document}
%%%%%%%%%%%%%%%%%%%%%%%%%%%%%%%%%%%%%%%%%%%%%%%%%%%%
%% common usage
\newcommand{\ks}{K_{S}^{0}}
\newcommand{\EP}{e^{+}}
\newcommand{\EM}{e^{-}}
\newcommand{\epm}{e^{\pm}}
\newcommand{\vpho}{\gamma^{\ast}}
\newcommand{\qqbar}{q\bar{q}}

%%%%%%%%%%%%%%%%%%%%%%%%%%%%%%%%%%%%%%%%%%%%%%%%%%%%
%% event selection
\newcommand{\ee}{e^{+}e^{-}}
\newcommand{\mm}{\mu^{+}\mu^{-}}
\newcommand{\alfs}{\alpha_{s}}
\newcommand{\alfmz}{\alpha(M_{Z}^{2})}
\newcommand{\amu}{a_{\mu}}
\newcommand{\Lam}{\Lambda_{c}}
\newcommand{\lam}{\Lambda_{c}^{+}}
\newcommand{\lambar}{\bar{\Lambda}_{c}^{-}}
\newcommand{\Lambdac}{\Lambda_{c}}
\newcommand{\mbc}{M_{BC}}
\newcommand{\dele}{\Delta E}
\newcommand{\ebm}{E_{\textmd{beam}}}
\newcommand{\ecm}{E_{\textmd{c.m.}}}
\newcommand{\pbm}{p_{\textmd{beam}}}
\newcommand{\MuMu}{\mu\mu}
\newcommand{\mumu}{\mu\mu}
\newcommand{\tata}{\tau^{+}\tau^{-}}
\newcommand{\pipi}{\pi^{+}\pi^{-}}
\newcommand{\gaga}{\gamma\gamma}
\newcommand{\twopho}{\ee+X}
\newcommand{\sqs}{\sqrt{s}}
\newcommand{\sqsp}{\sqrt{s^{\prime}}}
\newcommand{\da}{\Delta\alpha}
\newcommand{\das}{\Delta\alpha(s)}
\newcommand{\dimu}{\ee \ra \mumu}
\newcommand{\dedx}{\textmd{d}E/\textmd{d}x}
\newcommand{\chip}{\chi_{\textmd{Prob}}}
\newcommand{\chiP}{\chi_{p}}
\newcommand{\evz}{V_{z}^{\textmd{evt}}}
\newcommand{\evzloose}{V_{z,\textmd{loose}}^{\textmd{evt}}}
\newcommand{\avz}{V_{z}^{\textmd{ave}}}
\newcommand{\Ngd}{N_{\textmd{good}}}
\newcommand{\Ncru}{N_{\textmd{crude}}}
\newcommand{\pio}{\pi^{0}}
\newcommand{\rpid}{r_{\textmd{PID}}}

%%%%%%%%%%%%%%%%%%%%%%%%%%%%%%%%%%%%%%%%%%%%%%%%%%%%
%% simulation of signals
\newcommand{\Nhxobs}{N_{h+X}^{\textmd{obs}}}
\newcommand{\Nhobs}{N_{h}^{\textmd{obs}}}
\newcommand{\Npioxobs}{N_{\pi^{0}+X}^{\textmd{obs}}}
\newcommand{\Nksxobs}{N_{\ks+X}^{\textmd{obs}}}
\newcommand{\Npioobs}{N_{\pi^{0}}^{\textmd{obs}}}
\newcommand{\Nksobs}{N_{\ks}^{\textmd{obs}}}
\newcommand{\Nhxtru}{N_{h+X}^{\textmd{tru}}}
\newcommand{\Nhtru}{N_{h}^{\textmd{tru}}}
\newcommand{\Npiotru}{N_{\pi^{0}}^{\textmd{tru}}}
\newcommand{\Nkstru}{N_{\ks}^{\textmd{tru}}}
\newcommand{\Nbarhxobs}{\bar{N}_{h+X}^{\textmd{obs}}}
\newcommand{\Nbarhobs}{\bar{N}_{h}^{\textmd{obs}}}
\newcommand{\Nbarpioobs}{\bar{N}_{\pi^{0}}^{\textmd{obs}}}
\newcommand{\Nbarksobs}{\bar{N}_{\ks}^{\textmd{obs}}}
\newcommand{\Nbarhxtru}{\bar{N}_{h+X}^{\textmd{tru}}}
\newcommand{\Nbarhtru}{\bar{N}_{h}^{\textmd{tru}}}
\newcommand{\Nbarpiotru}{\bar{N}_{\pi^{0}}^{\textmd{tru}}}
\newcommand{\Nbarkstru}{\bar{N}_{\ks}^{\textmd{tru}}}

\newcommand{\Nhadtot}{N_{\textmd{had}}^{\textmd{tot}}}
\newcommand{\Nhadobs}{N_{\textmd{had}}^{\textmd{obs}}}
\newcommand{\Nbarhadobs}{\bar{N}_{\textmd{had}}^{\textmd{obs}}}
\newcommand{\Nhadtru}{N_{\textmd{had}}^{\textmd{tru}}}
\newcommand{\Nbarhadtru}{\bar{N}_{\textmd{had}}^{\textmd{tru}}}
\newcommand{\Nhadphy}{N_{\textmd{had}}}

\newcommand{\cshadobs}{\sigma_{\textmd{had}}^{\textmd{obs}}}
\newcommand{\effhad}{\vap_{\textmd{had}}}
\newcommand{\efftrg}{\vap_{\textmd{trig}}}
\newcommand{\lint}{\mathcal{L}_{\textmd{int}}}
\newcommand{\Nbkg}{N_{\textmd{bkg}}}
\newcommand{\NbkgTot}{N_{\textrm{bkg}}^{\textrm{Tot}}}
\newcommand{\csbkg}{\sigma_{\textmd{bkg}}}
\newcommand{\Nmcsur}{N_{\textmd{MC}}^{\textmd{sur}}}
\newcommand{\Nmcsurori}{N_{\textmd{MC}}^{\textmd{sur,nom.}}}
\newcommand{\Nmcsurwtd}{N_{\textmd{MC}}^{\textmd{sur,wtd.}}}
\newcommand{\Nmcgen}{N_{\textmd{MC}}^{\textmd{gen}}}
\newcommand{\vap}{\varepsilon}
\newcommand{\chisq}{\chi^{2}}
\newcommand{\cshadphy}{\sigma_{\textmd{had}}^{\textmd{phy}}}
\newcommand{\cshadtot}{\sigma_{\textmd{had}}^{\textmd{tot}}}
\newcommand{\cshadborn}{\sigma_{\textmd{had}}^{0}}
\newcommand{\cshadborncon}{\sigma_{\textmd{con}}^{0}}
\newcommand{\cshadbornres}{\sigma_{\textmd{res}}^{0}}
\newcommand{\csdimuborn}{\sigma_{\mu\mu}^{0}}
\newcommand{\rpqcd}{R_{\textmd{pQCD}}}
\newcommand{\Nprod}{N_{\textmd{prod}}}
\newcommand{\Nhadnet}{N_{\textrm{had}}^{\textrm{net}}}
\newcommand{\Delrel}{\Delta_{\textrm{rel}}}

%%%%%%%%%%%%%%%%%%%%%%%%%%%%%%%%%%%%%%%%%%%%%%%%%%%%
%% exclusive checks
\newcommand{\fourpionchg}{\pipi\pipi}
\newcommand{\fourpionneu}{\pipi\pi^{0}\pi^{0}}
\newcommand{\sixpionchg}{3(\pipi)}
\newcommand{\thrpionneu}{\pipi\pi^{0}}
\newcommand{\twopionchg}{\pipi}

\newcommand{\Nsurnpion}{N_{\textmd{sur}}^{n\pi}}
\newcommand{\Ngennpion}{N_{\textmd{gen}}^{n\pi}}
\newcommand{\Ngentot}{N_{\textmd{gen}}^{\textmd{tot}}}
\newcommand{\effincnpion}{\vap_{n\pi}^{\textmd{inc}}}
\newcommand{\effincnpionp}{\vap_{n\pi}^{\textmd{inc},\prime}}
\newcommand{\effincnonnpion}{\vap_{\textmd{non}-n\pi}^{\textmd{inc}}}
\newcommand{\effexcnpion}{\vap_{n\pi}^{\textmd{exc}}}
\newcommand{\fracnpion}{f_{n\pi}}
\newcommand{\fracnpionp}{f_{n\pi}^{\prime}}
\newcommand{\fracnonnpion}{f_{\textmd{non}-n\pi}}

\newcommand{\Nsurtwopion}{N_{\textmd{sur}}^{2\pi}}
\newcommand{\Ngentwopion}{N_{\textmd{gen}}^{2\pi}}
\newcommand{\effinctwopion}{\vap_{2\pi}^{\textmd{inc}}}
\newcommand{\effinctwopionp}{\vap_{2\pi}^{\textmd{inc},\prime}}
\newcommand{\effincnontwopion}{\vap_{\textmd{non}-2\pi}^{\textmd{inc}}}
\newcommand{\effexctwopion}{\vap_{2\pi}^{\textmd{exc}}}
\newcommand{\fractwopion}{f_{2\pi}}
\newcommand{\fractwopionp}{f_{2\pi}^{\prime}}
\newcommand{\fracnontwopion}{f_{\textmd{non}-2\pi}}

\newcommand{\Nsurthrpion}{N_{\textmd{sur}}^{3\pi}}
\newcommand{\Ngenthrpion}{N_{\textmd{gen}}^{3\pi}}
\newcommand{\effincthrpion}{\vap_{3\pi}^{\textmd{inc}}}
\newcommand{\effincthrpionp}{\vap_{3\pi}^{\textmd{inc},\prime}}
\newcommand{\effincnonthrpion}{\vap_{\textmd{non}-3\pi}^{\textmd{inc}}}
\newcommand{\effexcthrpion}{\vap_{3\pi}^{\textmd{exc}}}
\newcommand{\fracthrpion}{f_{3\pi}}
\newcommand{\fracthrpionp}{f_{3\pi}^{\prime}}
\newcommand{\fracnonthrpion}{f_{\textmd{non}-3\pi}}

\newcommand{\Nsurfourpion}{N_{\textmd{sur}}^{4\pi}}
\newcommand{\Ngenfourpion}{N_{\textmd{gen}}^{4\pi}}
\newcommand{\effincfourpion}{\vap_{4\pi}^{\textmd{inc}}}
\newcommand{\effincfourpionp}{\vap_{4\pi}^{\textmd{inc},\prime}}
\newcommand{\effincnonfourpion}{\vap_{\textmd{non}-4\pi}^{\textmd{inc}}}
\newcommand{\effexcfourpion}{\vap_{4\pi}^{\textmd{exc}}}
\newcommand{\fracfourpion}{f_{4\pi}}
\newcommand{\fracfourpionp}{f_{4\pi}^{\prime}}
\newcommand{\fracnonfourpion}{f_{\textmd{non}-4\pi}}

\newcommand{\Npionprod}{N_{\textmd{prod}}^{4\pi}}
\newcommand{\Ndatasur}{N_{\textmd{data}}^{\textmd{sur}}}
\newcommand{\Nobspion}{N_{\textmd{obs}}^{4\pi}}
\newcommand{\Nhadprod}{N_{\textmd{prod}}^{\textmd{had}}}
\newcommand{\sigmaobs}{\sigma_{\textmd{obs}}}
\newcommand{\effhadp}{\vap_{\textmd{had}}^{\prime}}

\newcommand{\effpion}{\vap_{4\pi}}
\newcommand{\effexcpion}{\vap_{4\pi}^{\textmd{exc}}}
\newcommand{\effincpion}{\vap_{4\pi}^{\textmd{inc}}}
\newcommand{\effincpionI}{\vap_{4\pi}^{\textmd{inc},1}}
\newcommand{\effincpionII}{\vap_{4\pi}^{\textmd{inc},2}}
\newcommand{\effincpionp}{\vap_{4\pi}^{\textmd{inc},\prime}}
\newcommand{\effincremain}{\vap_{\textmd{non}-n\pi}^{\textmd{inc}}}

\newcommand{\fracpion}{f_{4\pi}}
\newcommand{\fracnonpion}{f_{\textmd{non}-4\pi}}
\newcommand{\fracnonpionp}{f_{\textmd{non}-4\pi}^{\prime}}
\newcommand{\fracpionII}{f_{4\pi}^{2}}
\newcommand{\fracpionp}{f_{4\pi}^{\prime}}
\newcommand{\reladiff}{\Delta_{\textmd{rel}}}

\newcommand{\Nsursixpion}{N_{\textmd{sur}}^{6\pi}}
\newcommand{\Ngensixpion}{N_{\textmd{gen}}^{6\pi}}
\newcommand{\effincsixpion}{\vap_{6\pi}^{\textmd{inc}}}
\newcommand{\fracsixpion}{f_{6\pi}}

\newcommand{\etot}{E_{\textmd{tot}}}
\newcommand{\ptot}{p_{\textmd{tot}}}
\newcommand{\plab}{p_{\textmd{Lab}}}
\newcommand{\mpiOI}{M(\pi^{0}_{1})}
\newcommand{\mpiOII}{M(\pi^{0}_{2})}

%%%%%%%%%%%%%%%%%%%%%%%%%%%%%%%%%%%%%%%%%%%%%%%%%%%%
%% ISR corrections
\newcommand{\widtheeoi}{\varGamma^{\textmd{ee}}_{0,i}}
\newcommand{\widtheeoj}{\varGamma^{\textmd{ee}}_{0,j}}
\newcommand{\widtheeo}{\varGamma^{\textmd{ee}}_{0}}
\newcommand{\widthee}{\varGamma^{\textmd{ee}}}
\newcommand{\widtheeexpi}{\varGamma^{\textmd{ee}}_{\textmd{exp},i}}
\newcommand{\widtheeexp}{\varGamma^{\textmd{ee}}_{\textmd{exp}}}
\newcommand{\widthtoti}{\varGamma^{\textmd{tot}}_{i}}
\newcommand{\widthtot}{\varGamma^{\textmd{tot}}}

\newcommand{\vpqed}{\Pi_{\textmd{QED}}}
\newcommand{\vpqcd}{\Pi_{\textmd{QCD}}}
\newcommand{\vpcon}{\Pi_{\textmd{con}}}
\newcommand{\vpres}{\Pi_{\textmd{res}}}
\newcommand{\vpo}{\Pi_{0}}
\newcommand{\rcon}{R_{\textmd{con}}}
\newcommand{\rres}{R_{\textmd{res}}}
\newcommand{\rexp}{R_{\textmd{exp}}}

\newcommand{\delvert}{\delta_{\textmd{vert}}}
\newcommand{\delvp}{\delta_{\textmd{vac}}}
\newcommand{\delbrem}{\delta_{\gamma}}
\newcommand{\delobs}{\delta_{\textmd{obs}}}
\newcommand{\radiatorsf}{F_{\textmd{SF}}}
\newcommand{\radiatorfd}{F_{\textmd{FD}}}
\newcommand{\DelFD}{\Delta_{\textmd{FD}}}
\newcommand{\DelFDCal}{\Delta_{\textmd{cal}}}
\newcommand{\DelFDcs}{\Delta_{\sigma}}
\newcommand{\DelFDvp}{\Delta_{\textmd{vp}}}

%%%%%%%%%%%%%%%%%%%%%%%%%%%%%%%%%%%%%%%%%%%%%%%%%%%%
%% more comparisons
\newcommand{\costh}{\cos\theta}
\newcommand{\costhIprg}{\cos\theta_{\textmd{1prg}}}
\newcommand{\costhIIprg}{\cos\theta_{\textmd{2prg}}}
\newcommand{\costhIIIprg}{\cos\theta_{\textmd{3prg}}}
\newcommand{\costhIVprg}{\cos\theta_{\textmd{4prg}}}
\newcommand{\costhrestprg}{\cos\theta_{\textmd{restprg}}}
\newcommand{\emce}{E^{\textmd{ctrk.}}_{\textmd{emc}}}
\newcommand{\emceIprg}{E^{\textmd{ctrk.}}_{\textmd{emc,1prg}}}
\newcommand{\emceIIprg}{E^{\textmd{ctrk.}}_{\textmd{emc,2prg}}}
\newcommand{\emceIIIprg}{E^{\textmd{ctrk.}}_{\textmd{emc,3prg}}}
\newcommand{\emceIVprg}{E^{\textmd{ctrk.}}_{\textmd{emc,4prg}}}
\newcommand{\emcerestprg}{E^{\textmd{ctrk.}}_{\textmd{emc,restprg}}}
\newcommand{\isocosth}{\cos\theta_{\textmd{iso}}}
\newcommand{\isocosthIprg}{\cos\theta_{\textmd{iso,1prg}}}
\newcommand{\isocosthIIprg}{\cos\theta_{\textmd{iso,2prg}}}
\newcommand{\isocosthIIIprg}{\cos\theta_{\textmd{iso,3prg}}}
\newcommand{\isocosthIVprg}{\cos\theta_{\textmd{iso,4prg}}}
\newcommand{\isocosthrestprg}{\cos\theta_{\textmd{iso,restprg}}}
\newcommand{\eop}{E/P}
\newcommand{\eopIprg}{E/P_{\textmd{1prg}}}
\newcommand{\eopIIprg}{E/P_{\textmd{2prg}}}
\newcommand{\eopIIIprg}{E/P_{\textmd{3prg}}}
\newcommand{\eopIVprg}{E/P_{\textmd{4prg}}}
\newcommand{\eoprestprg}{E/P_{\textmd{restprg}}}
\newcommand{\nisogam}{N_{\textmd{isogam}}}
\newcommand{\nisogamIprg}{N_{\textmd{isogam,1prg}}}
\newcommand{\nisogamIIprg}{N_{\textmd{isogam,2prg}}}
\newcommand{\nisogamIIIprg}{N_{\textmd{isogam,3prg}}}
\newcommand{\nisogamIVprg}{N_{\textmd{isogam,4prg}}}
\newcommand{\nisogamrestprg}{N_{\textmd{isogam,restprg}}}
\newcommand{\ptrk}{p_{\textmd{ctrk}}}
\newcommand{\pIprg}{p^{\textmd{ctrk}}_{\textmd{1prg}}}
\newcommand{\pIIprg}{p^{\textmd{ctrk}}_{\textmd{2prg}}}
\newcommand{\pIIIprg}{p^{\textmd{ctrk}}_{\textmd{3prg}}}
\newcommand{\pIVprg}{p^{\textmd{ctrk}}_{\textmd{4prg}}}
\newcommand{\prestprg}{p^{\textmd{ctrk}}_{\textmd{restprg}}}
\newcommand{\tote}{E_{\textmd{vis.}}}
\newcommand{\totevte}{E_{\textmd{tot.}}}
\newcommand{\totevteIprg}{E_{\textmd{tot.}}^{\textmd{1prg}}}
\newcommand{\balanceIprg}{\textmd{Balance}}
\newcommand{\ngamma}{N_{\gamma}}
\newcommand{\ngammaIprg}{N_{\gamma,\textmd{1prg}}}
\newcommand{\ngammaIIprg}{N_{\gamma,\textmd{2prg}}}
\newcommand{\ngammaIIIprg}{N_{\gamma,\textmd{3prg}}}
\newcommand{\ngammaIVprg}{N_{\gamma,\textmd{4prg}}}
\newcommand{\ngammarestprg}{N_{\gamma,\textmd{restprg}}}
\newcommand{\ngoodwt}{N_{\textmd{good}}^{\textmd{Wt}}}
\newcommand{\ngood}{N_{\textmd{good}}}
\newcommand{\npiO}{N_{\pi^{0}}}
\newcommand{\npiOIprg}{N_{\pi^{0}}^{\textmd{1prg}}}
\newcommand{\npiOIIprg}{N_{\pi^{0}}^{\textmd{2prg}}}
\newcommand{\npp}{N_{p}}
\newcommand{\npm}{N_{\bar{p}}}
\newcommand{\nkp}{N_{K^{+}}}
\newcommand{\nkm}{N_{K^{-}}}
\newcommand{\npip}{N_{\pi^{+}}}
\newcommand{\npim}{N_{\pi^{-}}}
\newcommand{\ppp}{P(p^{+})}
\newcommand{\ppm}{P(\bar{p}^{-})}
\newcommand{\pkp}{p(K^{+})}
\newcommand{\pkm}{p(K^{-})}
\newcommand{\ppip}{P(\pi^{+})}
\newcommand{\ppim}{P(\pi^{-})}
\newcommand{\ppiO}{P(\pi^{0})}
\newcommand{\mpiO}{M(\pi^{0})}
\newcommand{\mks}{M(K^{0}_{s})}
\newcommand{\pks}{p_{K^{0}_{s}}}
\newcommand{\mphi}{M(\phi)}
\newcommand{\pphi}{p_{\phi}}
\newcommand{\mIIgam}{M(\gamma\gamma)}
\newcommand{\mIIgamIprg}{M(\gamma\gamma)^{\textmd{1prg}}}
\newcommand{\pIIgam}{p_{\gamma\gamma}}
\newcommand{\mlambda}{M(\Lambda)}
\newcommand{\plambda}{p_{\Lambda}}
\newcommand{\mdO}{M(D^{0})}
\newcommand{\pdO}{p_{D^{0}}}
\newcommand{\mdstarO}{M(D^{\ast 0})}
\newcommand{\pdstarO}{p_{D^{\ast 0}}}
\newcommand{\mdp}{M(D^{\pm})}
\newcommand{\pdp}{p_{D^{\pm}}}
\newcommand{\mdstarp}{M(D^{\ast\pm})}
\newcommand{\pdstarp}{p_{D^{\ast\pm}}}
\newcommand{\mds}{M(D_{s}^{\pm})}
\newcommand{\pds}{p_{D_{s}^{\pm}}}
\newcommand{\mdstars}{M(D_{s}^{\ast\pm})}
\newcommand{\pdstars}{p_{D_{s}^{\ast\pm}}}
\newcommand{\Vr}{V_{r}}
\newcommand{\Vz}{V_{z}}

\newcommand{\gev}{\mathrm{GeV}}
\newcommand{\mev}{\mathrm{MeV}}
\newcommand{\mevcc}{\mathrm{MeV}/c^{2}}
\newcommand{\gevc}{\mathrm{GeV}/c}
\newcommand{\gevcc}{\mathrm{GeV}/c^2}

%%%%%%%%%%%%%%%%%%%%%%%%%%%%%%%%%%%%%%%%%%%%%%%%%%%%
%unfolding
\newcommand{\nchg}{N_{\textmd{chg}}}
\newcommand{\eff}{\vap}

%%%%%%%%%%%%%%%%%%%%%%%%%%%%%%%%%%%%%%%%%%%%%%%%%%%%
%% other usages
\newcommand{\critecm}{1.780}

\newcommand{\ENERGYAT}{4575.5}
\newcommand{\ENERGYBT}{4575.5}
\newcommand{\ENERGYCT}{4575.5}
\newcommand{\ENERGYDT}{4575.5}
\newcommand{\ksdecay}{\ks\ra\pi^{+}\pi^{-}}
\newcommand{\phidecay}{\phi\ra K^{+}K^{-}}
\newcommand{\piOdecay}{\pi^{0}\ra\gamma\gamma}
\newcommand{\Lambdadecay}{\Lambda\ra p\pi^{-}}
\newcommand{\DOdecay}{D^{0}\ra K^{-}\pi^{+}}
\newcommand{\DStarOdecay}{D^{\ast0}\ra D^{0}\pi^{0}}
\newcommand{\Dpdecay}{D^{+}\ra K^{+}\pi^{+}\pi^{-}}
\newcommand{\DStarpdecay}{D^{\ast+}\ra D^{0}\pi^{+}}
\newcommand{\Dsdecay}{D^{+}_{s}\ra K^{+}K^{-}\pi^{+}}
\newcommand{\DStarsdecay}{D^{\ast+}_{s}\ra D^{+}_{s}\gamma}

%%%%%%%%%%%%%%%%%%%%%%%%%%%%%%%%%%%%%%%%%%%%%%%%%%%%%%%%%%%%%%%%%%%%%%%%%%%%%%%%%%%%%%%%%%%%%%%%%%%%%%%%%%%%%%%%%%%%%%%%%%%%%%%%%%%%%%%%%%%%%%%%%%%%%%%%%%%%%%%%%%%%%%%%%%%%%%%%%%%%%%%%%%%%%%

\title{\boldmath \textbf{Observation of an axial-vector state in the study of $\psip \to \phi \eta \etap $ decay} }
%% Saved at => 2022-07-21
\author{
\begin{small}
\begin{center}
M.~Ablikim$^{1}$, M.~N.~Achasov$^{4,b}$, P.~Adlarson$^{75}$, O.~Afedulidis$^{3}$, X.~C.~Ai$^{80}$, R.~Aliberti$^{35}$, A.~Amoroso$^{74A,74C}$, Q.~An$^{71,58}$, Y.~Bai$^{57}$, O.~Bakina$^{36}$
% ...
\\
\vspace{0.2cm}
(BESIII Collaboration)\\
\vspace{0.2cm} {\it
$^{1}$ Institute of High Energy Physics, Beijing 100049, People's Republic of China\\
% ...
$^{3}$ Bochum Ruhr-University, D-44780 Bochum, Germany\\
$^{4}$ Budker Institute of Nuclear Physics SB RAS (BINP), Novosibirsk 630090, Russia\\
% ...
$^{35}$ Johannes Gutenberg University of Mainz, Johann-Joachim-Becher-Weg 45, D-55099 Mainz, Germany\\
$^{36}$ Joint Institute for Nuclear Research, 141980 Dubna, Moscow region, Russia\\
% ...
$^{57}$ Southeast University, Nanjing 211100, People's Republic of China\\
$^{58}$ State Key Laboratory of Particle Detection and Electronics, Beijing 100049, Hefei 230026, People's Republic of China\\
% ...
$^{71}$ University of Science and Technology of China, Hefei 230026, People's Republic of China\\
% ...
$^{74}$ University of Turin and INFN, (A)University of Turin, I-10125, Turin, Italy; (B)University of Eastern Piedmont, I-15121, Alessandria, Italy; (C)INFN, I-10125, Turin, Italy\\
$^{75}$ Uppsala University, Box 516, SE-75120 Uppsala, Sweden\\
% ...
$^{80}$ Zhengzhou University, Zhengzhou 450001, People's Republic of China\\
\vspace{0.2cm}
% ...
$^{b}$ Also at the Novosibirsk State University, Novosibirsk, 630090, Russia\\
% ...
}
\end{center}
\vspace{0.4cm}
\vspace{0.4cm}
\end{small}
}

\noaffiliation{}

\date{\today}

\begin{abstract}

Using (2712.4~$\pm$~14.3)$\times 10^{6}$ $\psi(3686)$ events collected with the BESIII detector at BEPCII, a partial wave analysis of the decay $\psip \to \phi \eta \etap $ is performed with the covariant tensor approach. An axial-vector state with a mass near 2.3\,$\gevcc$ is observed for the first time.  Its mass and width are measured to be 2316 $\pm 9_{\mathrm{stat}} \pm 30_{\mathrm{syst}}\,\mevcc$ and 89 $\pm 15_{\mathrm{stat}} \pm 26_{\mathrm{syst}}\,\mev$, respectively. The product branching fractions of $\mathcal{B}(\psi(3686) \to X(2300) \eta') \mathcal{B}(X(2300)\to \phi \eta)$ and $\mathcal{B}(\psip \to X(2300) \eta)\mathcal{B}(X(2300)\to \phi \etap)$ are determined to be (4.8 $\pm 1.3_{\mathrm{stat}} \pm 0.7_{\mathrm{syst}})\times 10^{-6}$ and (2.2 $\pm 0.7_{\mathrm{stat}} \pm 0.7_{\mathrm{syst}})\times 10^{-6}$, respectively. The branching fraction $\mathcal{B}(\psi(3686) \to \phi \eta \eta')$ is measured for the first time to be (3.14$\pm0.17_{\mathrm{stat}}\pm0.24_{\mathrm{syst}})\times10^{-5}$.
  The first uncertainties are statistical and the second are systematic.

\end{abstract}

\maketitle

%%%%%%%%%%%%%%%%%%%%%%%%%%%%%%%%%%%%%%%%%%%%%%%%%%%%%%%%%%%%%%%%%%%%%%%%%%%%%%%%%%

{%\color{red}
Strangeonium ($s\bar{s}$) states can provide insight into the non-perturbative nature of Quantum Chromodynamics (QCD)~\cite{Godfrey:1985xj}. 
There are very few strangeonium resonances listed by the Particle Data Group (PDG)~\cite{ParticleDataGroup:2020ssz} that have been experimentally well confirmed and are widely accepted as pure $s\bar{s}$ states~\cite{Barnes:2002mu,Li:2020xzs}. 
For the spectrum of axial-vector strangeonium states, only the ground axial-vector state $h_1(1380)$~\cite{CrystalBarrel:1997kda,BESIII:2015vfb,BESIII:2018ede}
has been confrimed in PDG~\cite{ParticleDataGroup:2020ssz}, 
and excited states, e.g., $h_1(2P)$ and $h_1(3P)$, have been predicted by theory~\cite{Barnes:2002mu,Li:2020xzs,h13p_1,h13p_2,h13p_3,h13p_4,Wang:2019qyy} but have not been observed by experiment.
The systematic search for such states, and their subsequent investigation, will supplement our knowledge about the $h_1$ family and improve our understanding of the strangeonium spectrum~\cite{Chen:2015iqa}.
The $h_1$ family also provides interesting possibilities to study QCD in the non-perturbative regime via the mixing mechanism, since it is sensitive to the mixing of the flavor SU(3) singlet and octet states~\cite{Cheng:2011pb,Li:2005eq}.
Recently, BESIII observed the $1^{+-}$ state, which is a candidate for the $h_1(2P)$ state~\cite{BESIII:MEMO}. However, the $h_1(3P)$ still remains unobserved by experiments.

In addition, exotic states offer a distinctive setting for investigating the dynamics of strong interactions and the confinement mechanism. 
Recently, the $X(6900)$, $X(6600)$ and $X(7300)$ were observed by LHCb~\cite{LHCb:2020bwg}, CMS~\cite{CMS:2023owd} and ATLAS~\cite{ATLAS:2023bft} collaborations. These  resonances could be explained as fully-charm tetraquark states, $T_{cc\bar{c}\bar{c}}$~\cite{Galkin:2023wox,Agaev:2023gaq,wang:2020wrp,Bedolla:2019zwg,liu:2020eha,Giron:2020wpx}. 
%This intriguing discovery has sparked significant theoretical interest, with many researchers proposing that these new resonances could be attributed to a fully charmed tetraquark state, $T_{cc\bar{c}\bar{c}}$~\cite{Galkin:2023wox,Agaev:2023gaq,wang:2020wrp,Bedolla:2019zwg,liu:2020eha,Giron:2020wpx}. 
Analogous to the $T_{cc\bar{c}\bar{c}}$ system, 
it is conceivable that stable states such as stable fully-bottom tetraquark $T_{bb\bar{b}\bar{b}}$ and fully-strange tetraquark $T_{ss\bar{s}\bar{s}}$ states might exist. 
Just like the fully heavy tetraquarks, in the fully-strange tetraquarks 
there are no light meson exchanges, which are usually considered to 
be the dynamic mechanism for the formation of hadronic molecules. 
%The search for $T_{ss\bar{s}\bar{s}}$ states is instrumental in enhancing our comprehension of $T_{cc\bar{c}\bar{c}}$ states and helpful to construct a complete tetraquark spectrum.
Theory predicted the $T_{ss\bar{s}\bar{s}}$ spectrum, and proposed the highlighted decay modes such as $\phi\eta$ and $\phi\eta'$ as  crucial  channels to search the  $T_{ss\bar{s}\bar{s}}$ states through quark rearrangements~\cite{tetraquark:2021,Lu:2019ira,Cui:2019roq}.
%Theoretical predictions $T_{ss\bar{s}\bar{s}}$ 
%indicate that the decay modes such as $\phi\eta$ and $\phi\eta^{'}$ could serve as effective channels to investigate the  $T_{ss\bar{s}\bar{s}}$ states through quark rearrangements~\cite{tetraquark:2021,Lu:2019ira,Cui:2019roq}.
%the $T_{ss\bar{s}\bar{s}}$ could readily decay into $\phi\eta$ and $\phi\eta^{'}$ through quark rearrangements.
The BESIII experiment has collected the world's largest $\jpsi$ and $\psip$ data sets~\cite{BESIII:MEMO1,Numberjpsi}, and provides a unique opportunity to search $T_{ss\bar{s}\bar{s}}$ and $s\bar{s}$ states in the charmonium decays.
 
%In addition, strangeonium ($s\bar{s}$) states, which occupy an intermediate position between light and heavy quarkonia, can provide insight into the non-perturbative, low-energy regime of Quantum Chromodynamics (QCD), where the effective theory of heavy quarks is not suitable~\cite{Godfrey:1985xj}. 
}

Based on (2712.4~$\pm$~14.3) $\times 10^6$ $\psip$ events accumulated by the BESIII experiment, a partial wave analysis (PWA) is performed on the decay $\psip \to \phi \eta \etap$, where the $\phi$, $\eta$, and $\eta^\prime$ mesons are reconstructed via  $\phi\to K^+K^-$, $\eta\to \gamma\gamma$, and $\eta^\prime\to \gamma \pi^+\pi^-$, respectively.  This Letter presents an observation of a new axial-vector state around 2.3\,$\gevcc$ in the $\phi\eta$ and $\phi\etap$ invariant mass spectra.

The BESIII detector records symmetric $e^+e^-$ collisions  provided by the BEPCII storage ring~\cite{Yu:IPAC2016-TUYA01}
in the center-of-mass energy range from 2.0 to 4.95~GeV, with a peak luminosity of $1 \times 10^{33}\;\text{cm}^{-2}\text{s}^{-1}$ 
achieved at $\sqrt{s} = 3.77\;\text{GeV}$.
%which is of  the center-of-mass energy ($\sqrt{s}$) ranging from 2.0 to 4.95~$\gev$ and a peak luminosity of $1 \times 10^{33}\;\text{cm}^{-2}\text{s}^{-1}$ achieved at $\sqrt{s} = 3.77\;\text{GeV}$. 
The detailed description of the BESIII detector can be found in Ref.~\cite{Ablikim:2009aa}.
%The BESIII detector~\cite{Ablikim:2009aa} records symmetric $e^+e^-$ %collisions provided by the BEPCII storage ring~\cite{Yu:IPAC2016-%TUYA01} in the center-of-mass energy range from 2.0 to 4.95\,GeV,
%with a peak luminosity of $1 \times %10^{33}\;\text{cm}^{-2}\text{s}^{-1}$ achieved at $\sqrt{s} = 3.77\;\text{GeV}$. 
%The cylindrical core of the BESIII detector covers 93\% of the full solid angle and consists of a helium-based
% multilayer drift chamber~(MDC), a plastic scintillator time-of-flight system~(TOF), and a CsI(Tl) electromagnetic calorimeter~(EMC),
%which are all enclosed in a superconducting solenoidal magnet
%providing a 1.0~T (0.9~T in 2012) magnetic field. The solenoid is %supported by an
%octagonal flux-return yoke with resistive plate counter muon
%identification modules interleaved with steel. 
%The charged-particle momentum resolution at $1~{\rm GeV}/c$ is
%$0.5\%$, and the 
%${\rm d}E/{\rm d}x$
%resolution is $6\%$ for electrons
%from Bhabha scattering. The EMC measures photon energies with a
%resolution of $2.5\%$ ($5\%$) at $1$~GeV in the barrel (end cap)
%region. The time resolution in the TOF barrel region is 68~ps, while
%that in the end cap region is 110~ps. The end cap TOF
%system was upgraded in 2015 using multigap resistive plate chamber
%technology, providing a time resolution of 60~ps~\cite{etof}, which benefits 83\% of the data used in this analysis.
Simulated samples produced with {\sc geant4}-based~\cite{GEANT4:2002zbu} Monte Carlo (MC) software, which includes the geometric description of the BESIII detector~\cite{detvis} and the detector response, are used to determine detection efficiencies and to estimate backgrounds. The simulation models
the beam energy spread and initial state radiation in the $e^+e^-$ annihilation with the generator {\sc kkmc}~\cite{Jadach:2000ir,Jadach:1999vf}. An inclusive MC sample is generated to study the potential background with the production of the $\psip$ resonance simulated by the {\sc kkmc} generator~\cite{Jadach:2000ir,Jadach:1999vf}. The known decay modes are modeled with {\sc evtgen}~\cite{Lange:2001uf,Ping:2008zz} using branching fractions taken from the PDG~\cite{ParticleDataGroup:2020ssz}, and the remaining unknown charmonium decays are modeled with {\sc lundcharm}~\cite{Chen:2000tv,Yang:2014vra}. Final state radiation from charged final state particles is incorporated using {\sc photos}~\cite{Barberio:1993qi}.

Event candidates are required to have four charged tracks with zero net charge and at least three photons.
Charged tracks detected in the multilayer drift chamber~(MDC) are required to be within a polar angle ($\theta$) range of $|\rm{cos\theta}|<0.93$, where $\theta$ is defined with respect to the detector symmetry axis ($z$-axis), and their distance of closest approach to the interaction point must be less than 10\,cm along the $z$-axis and less than 1\,cm in the transverse plane.
 Information from the time-of-flight system~(TOF) and ${\rm d}E/{\rm d}x$ measurements is combined to form particle identification (PID) likelihoods for the $\pi$, $K$, and $p$ hypotheses. Each track is assigned as the particle type corresponding to the hypothesis with the highest PID likelihood. 
Exactly two oppositely charged kaons and pions are required in each event.
Photon candidates are identified using showers in the electromagnetic calorimeter~(EMC). The deposited energy of each shower is more than 25\,MeV in the barrel region ($|\cos \theta|< 0.80$) and more than 50\,MeV in the end cap region ($0.86 <|\cos \theta|< 0.92$). 
To exclude showers induced by charged particles, the angle between the position of each shower in the EMC and the closest extrapolated charged track is required to be larger than 10$^{\circ}$.
To suppress electronic noise and showers unrelated to the event, the difference between the EMC time and the event start time is required to be within (0, 700)\,ns. 

 \begin{figure*}[t]
 \centering
 \begin{overpic}[width=0.24\textwidth]{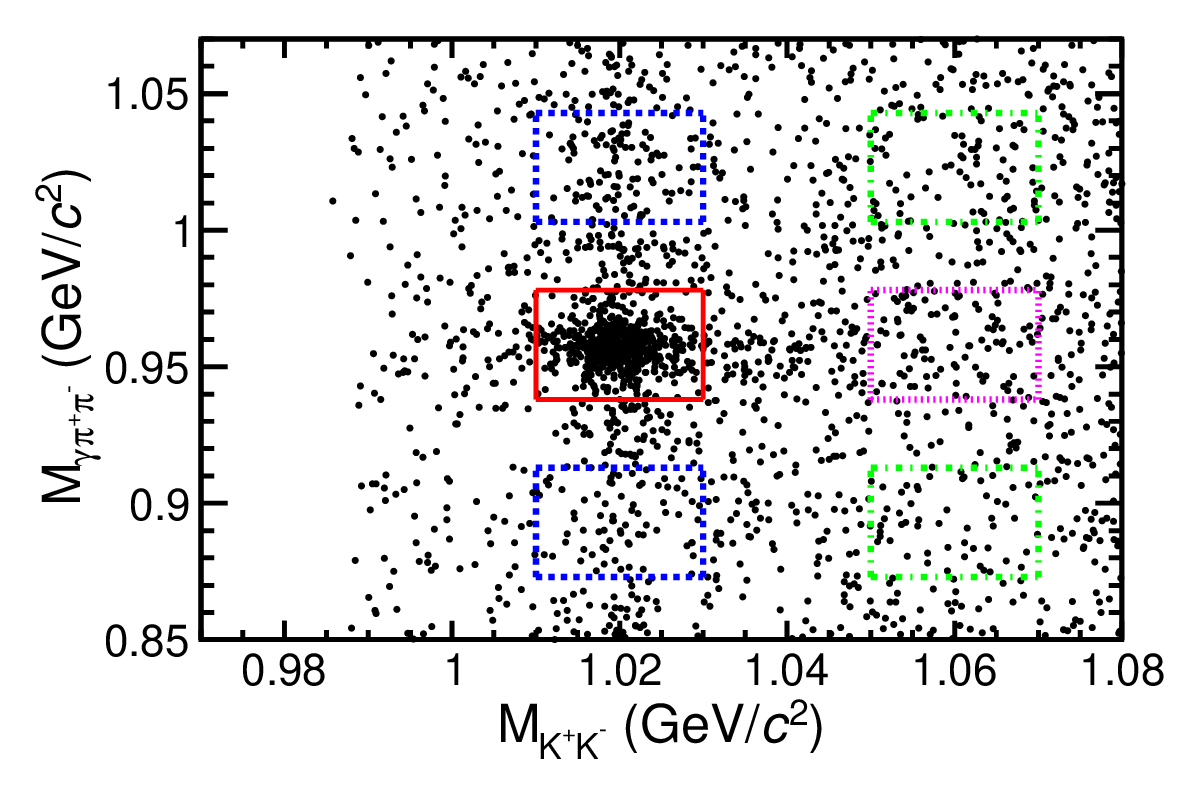}
 \put(20,55){ (a)}
 \end{overpic}
 \begin{overpic}[width=0.24\textwidth]{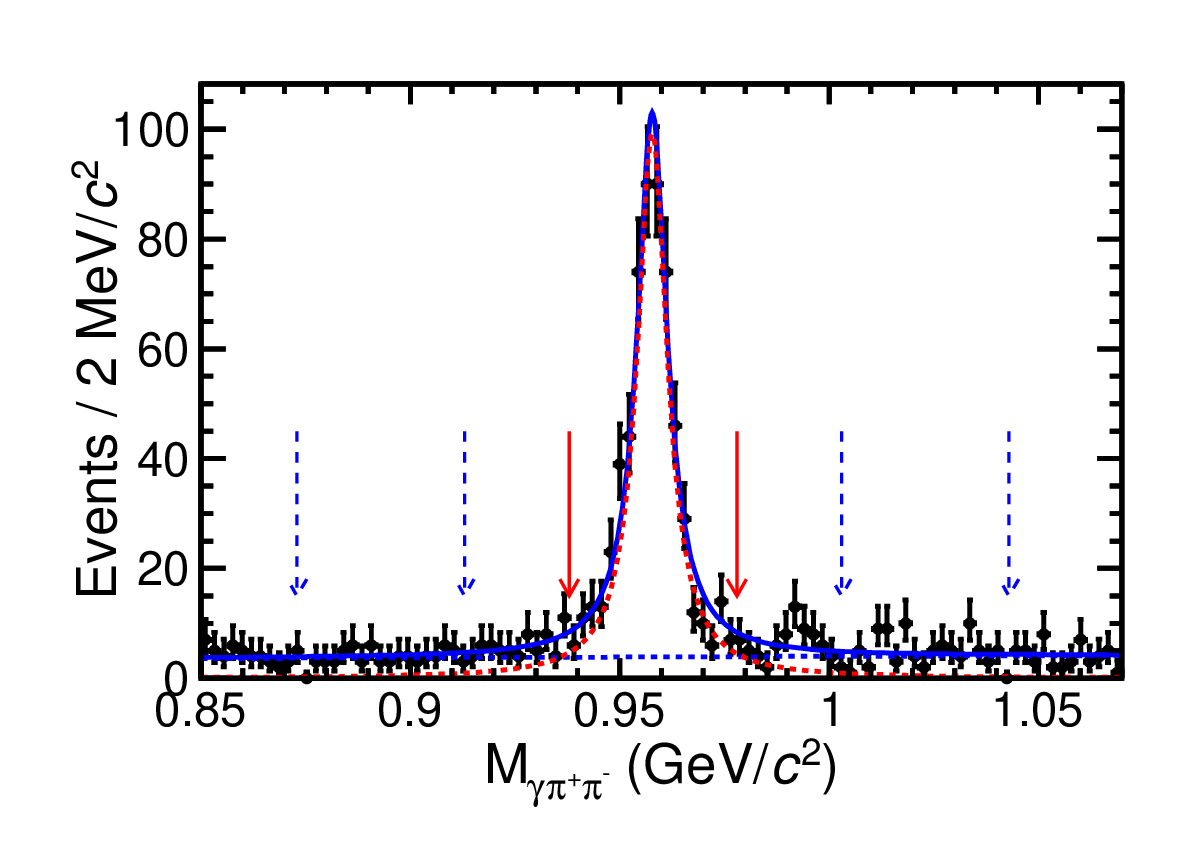}
 \put(20,55){ (b)}
 \end{overpic}
 \begin{overpic}[width=0.24\textwidth]{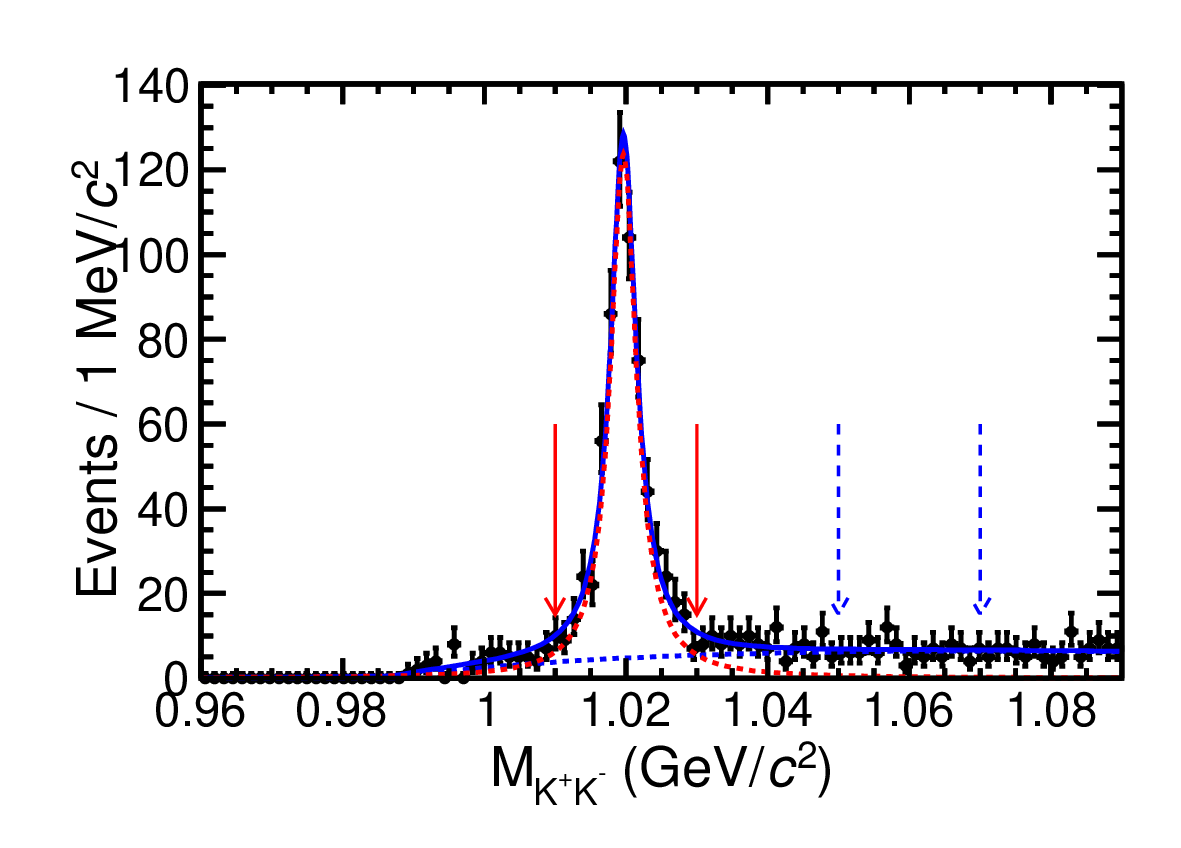}
 \put(20,55){ (c)}
 \end{overpic}
  \begin{overpic}[width=0.24\textwidth]{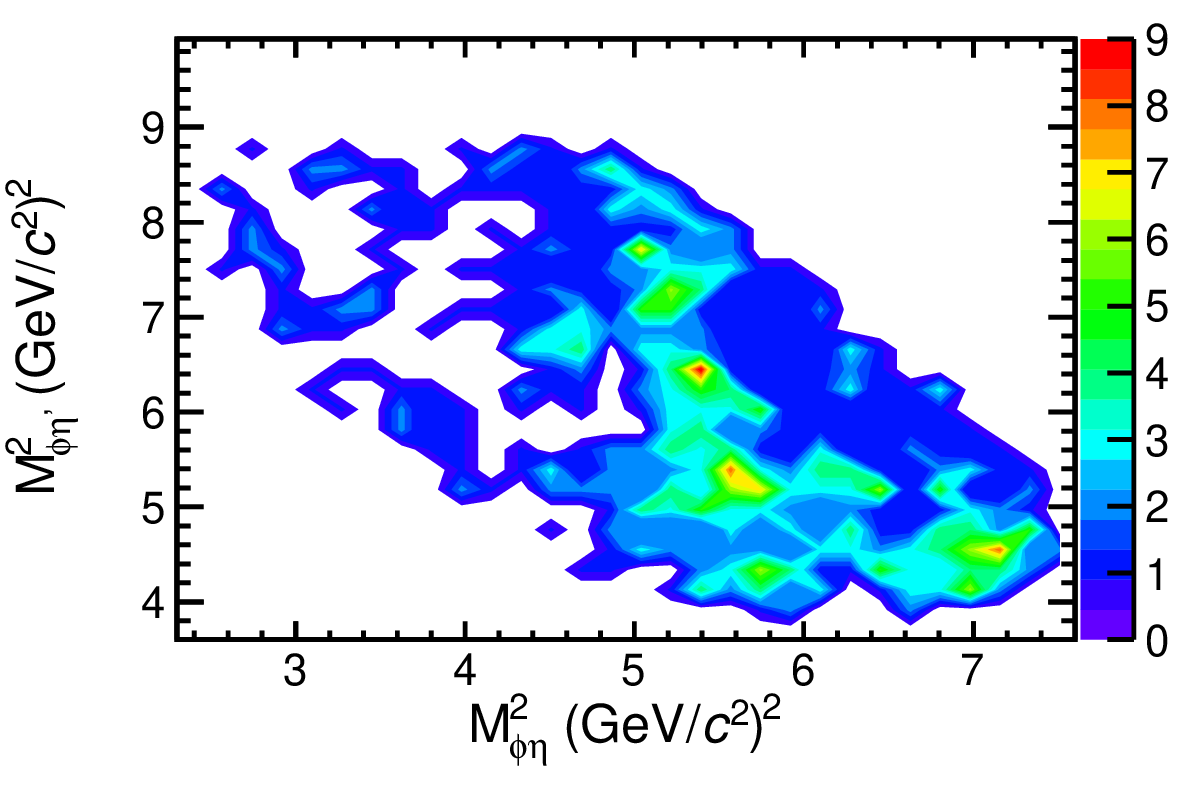}
 \put(70,55){ (d)}
 \end{overpic}
  \vspace*{-0.5cm}
 %\vspace*{100pt}
%\setlength{\abovecaptionskip}{1pt}
\caption{(a) Distribution of $M(\kpkm)$ versus $M(\gamma\pipi)$, where the red solid box shows the signal region and the dotted and dashed boxes are the sideband regions for the $\phi$ and $\etap$. (b)~Distribution of $M(\gamma\pipi)$ within the $\phi$ signal region. (c) Distribution of $M(\kpkm)$ within the $\etap$ signal region. (d)  Dalitz plot of $M^2(\phi\eta)$ versus $M^{2}(\phi\etap)$ in the signal regions. The solid arrows show the signal regions and the dashed arrows show the sideband regions.}
 \label{Fig:1}
\end{figure*}

 To improve the momentum resolution and to suppress background, a four-constraint (4C) kinematic fit imposing energy-momentum conservation with four degrees of freedom is carried out under the hypothesis of $\psip \to\kpkm\pipi\gamma\gamma\gamma$.
 For the events with more than three photons, the combination with the smallest $\chi_{4\rm{C}}^2$ of the 4C fit is retained, and $\chi^2_{4\rm{C}} <50$ is required. 
  The $\eta$ and $\etap$ candidates are reconstructed by minimizing $\chi^2_{\eta\etap} =(M_{\gamma_1\gamma_2}-m_{\eta})^2/\sigma_{\eta}^2+(M_{\gamma_3 \pi^{+} \pi^{-}}-m_{\etap})^2/\sigma_{\etap}^2$, where $m_{\eta}$ and $m_{\etap}$ are the nominal masses of the $\eta$ and $\etap$~\cite{ParticleDataGroup:2020ssz}, $M_{\gamma_1\gamma_2}$, and $M_{\gamma_3\pipi}$ are the invariant masses of the $\gamma_1\gamma_2$ and $\gamma_3\pipi$ combinations, and $\sigma_{\eta}$ and $\sigma_{\etap}$ are their corresponding resolutions determined from the signal MC simulations, respectively. 
 An event is accepted if $\eta$ and $\etap$ candidates satisfy $|M_{\gamma_1\gamma_2}-m_{\eta}| < 0.025~\gevcc$ and  $|M_{\gamma_3\pipi}-m_{\etap}| < 0.020~\gevcc$ requirements. The $\phi$ candidates are selected by requiring the $\kpkm$ invariant mass to satisfy $|M_{\kpkm}-m_{\phi}| < 0.010~\gevcc$, where $m_{\phi}$ is the $\phi$ nominal mass~\cite{ParticleDataGroup:2020ssz}.
 
 To suppress contamination from $\psip \to\kpkm\pipi\gamma\gamma\gamma\gamma$ and $\psip \to\kpkm\pipi\gamma\gamma$ decays, two additional 4C kinematic fits under the hypotheses of $\psip \to\kpkm\pipi\gamma\gamma\gamma\gamma$ and $\psip \to\kpkm\pipi\gamma\gamma$ are performed.
 The events are discarded if the corresponding $\chi^ {2}_{4\rm{C}}$ is less than the $\chi^ {2}_{4\rm{C}}$ for the signal hypothesis.
  To reject background from $\psip \to \gamma\chi_{cJ}$ with the subsequent decay $\chi_{cJ} \to \kpkm\pipi\gamma\gamma$ and background from $\psip \to \pipi J/\psi$ with the decay $J/\psi \to \gamma\gamma\gamma\kpkm$, the requirements $|M_{\kpkm\pipi\gamma\gamma}-M_{\chi_{cJ}}|$ $>0.010$\,$\gevcc$  and $|M_{\gamma\gamma\gamma\kpkm}-M_{J/\psi}|$ $>0.030$\,$\gevcc$ are applied. The requirement $|M_{\gamma\pi^{+}\pi^{-}\kpkm}-M_{J/\psi}|$\,$>0.030$ $\gevcc$ is used to remove the background of $\psip \to \eta J/\psi$ with $J/\psi \to \gamma\pi^{+}\pi^{-} K^{+} K^{-}$. To improve the momentum resolution, a 5C kinematic fit with an additional $\eta$ mass constraint is performed, and the resulting kinematic variables are used for further analysis.

After imposing all the selection criteria, the distribution of $M_{\kpkm}$ versus $M_{\gamma\pipi}$ is illustrated in Fig.~\ref{Fig:1}(a). A clear accumulation of candidate events for the decay $\psip \to \phi\eta\etap$ is observed. Figures~\ref{Fig:1}(b) and~\ref{Fig:1}(c) show the $M_{\gamma\pipi}$ and $M_{\kpkm}$ distributions. Potential backgrounds are studied using an inclusive MC sample of 2.747$\times10^{9}$ $\psip$ events. No significant peaking background is observed in the signal region. 
 Therefore, the two-dimensional sideband method is used to estimate the combinatorial backgrounds by combining the background events in the $\phi$ sideband region ($1.05\,\gevcc<M_{K^{+}K^{-}}<1.07\,\gevcc$) and the $\etap$ sideband region ($0.045\,\gevcc<|M_{\gamma\pi^{+}\pi^{-}}-M_{\etap}|<0.085\,\gevcc$), which are illustrated by the colored dashed boxes in Fig.~\ref{Fig:1}(a). 
 The normalization factors for the event yields in the two
sideband regions are obtained from the fits to the 
invariant mass spectra of $M_{\gamma\pipi}$ and $M_{\kpkm}$. The signal shapes are taken directly from signal MC simulation. The backgrounds are described with a first-order Chebychev polynomial function for $M_{\gamma\pipi}$ and an ARGUS function~\cite{ARGUS:1994rms} for $M_{\kpkm}$.
The selected data sample contains a total of 597$\pm$25 candidate events including 94$\pm$10 background events estimated from the sideband region and normalized to the signal region.

To investigate the continuum background from non-resonant $e^+e^-$ annihilation, the same selection criteria and sideband definition are applied in the analysis of the data sample taken at a center-of-mass energy of 
3.773~GeV corresponding to an integrated luminosity of 2.93~$\rm fb^{-1}$. After taking into account the integrated luminosity and the energy dependence of the cross sections, the continuum background contribution is estimated to be 23$\pm$7 events. By subtracting sideband and continuum background components from the total number of candidate events, 
the signal yield is determined to be $N_{\rm sig}=480\pm26$, where the uncertainty is statistical.

\begin{figure*}[htbp]
 \centering
 \begin{overpic}[width=0.32\textwidth]{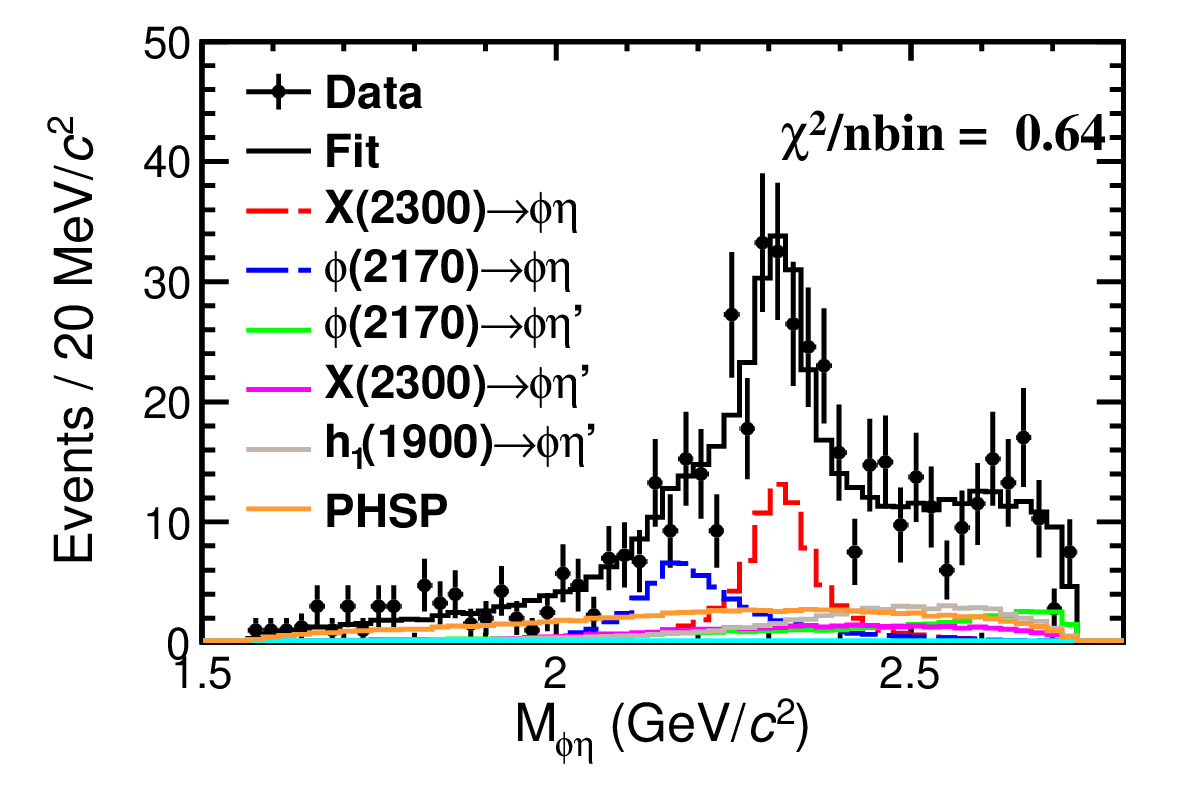}
 \put(80,48){ (a)}
 \end{overpic}
 \begin{overpic}[width=0.32\textwidth]{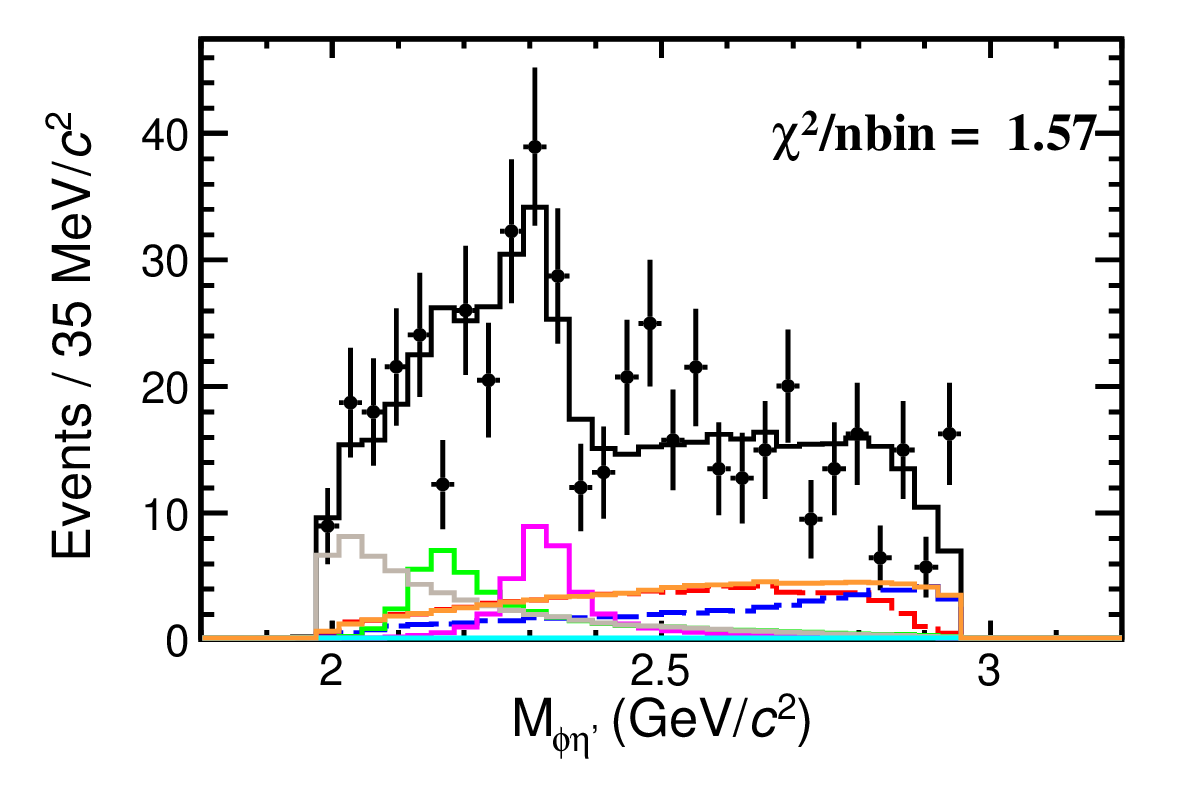}
 \put(80,48){ (b)}
 \end{overpic}
 \begin{overpic}[width=0.32\textwidth]{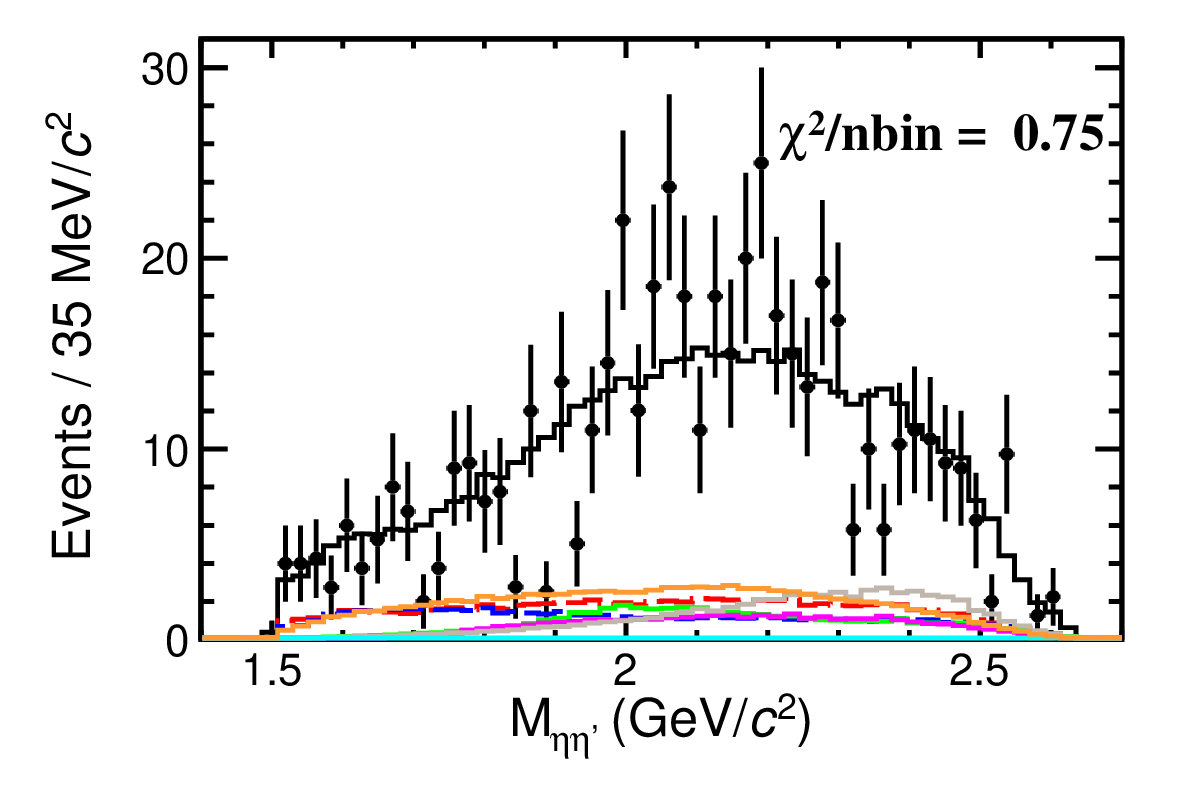}
 \put(80,48){ (c)}
 \end{overpic}

 \begin{overpic}[width=0.32\textwidth]{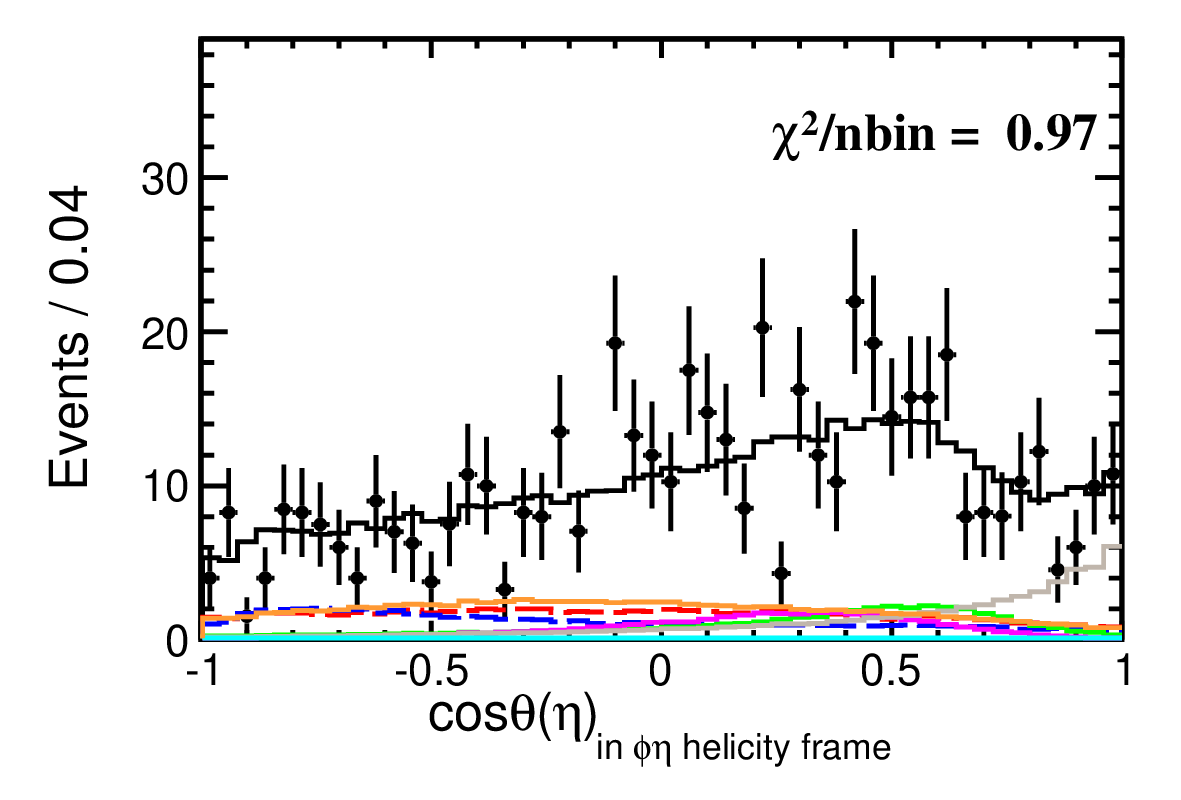}
\put(80,48){ (d)}
 \end{overpic}
 \begin{overpic}[width=0.32\textwidth]{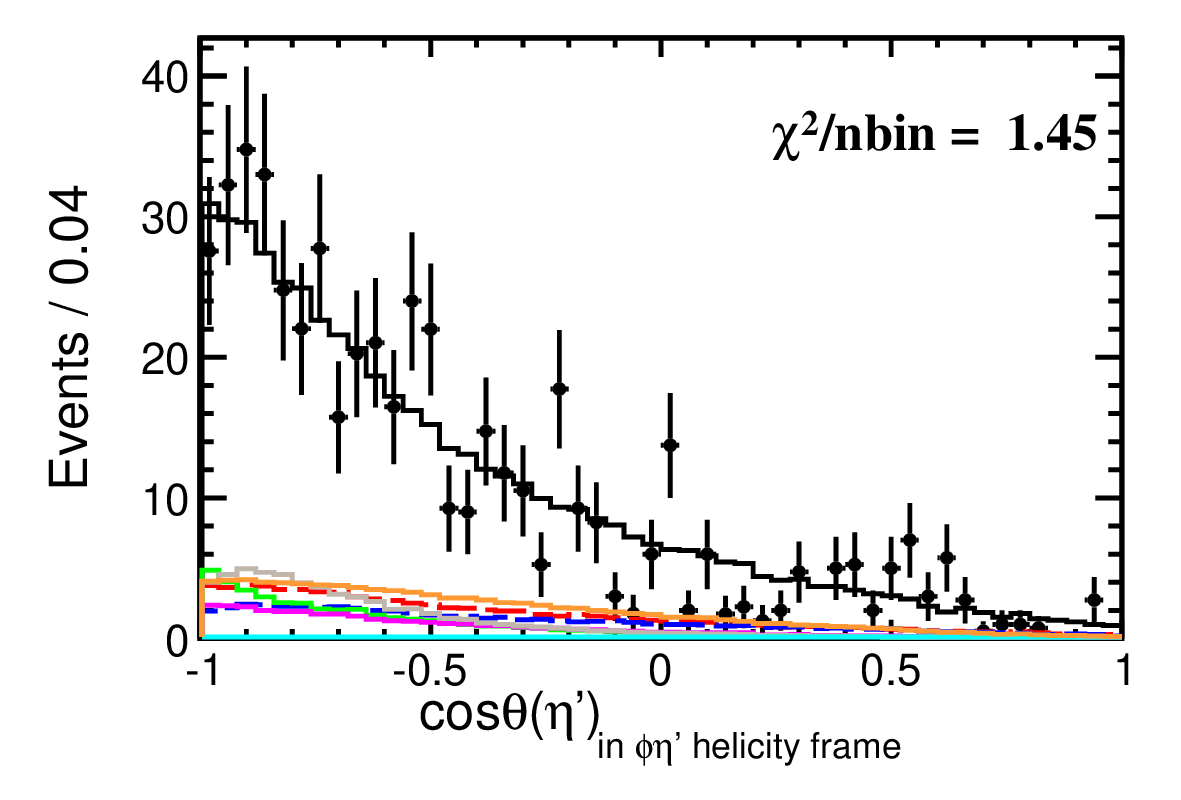}
 \put(80,48){ (e)}
 \end{overpic}
 \begin{overpic}[width=0.32\textwidth]{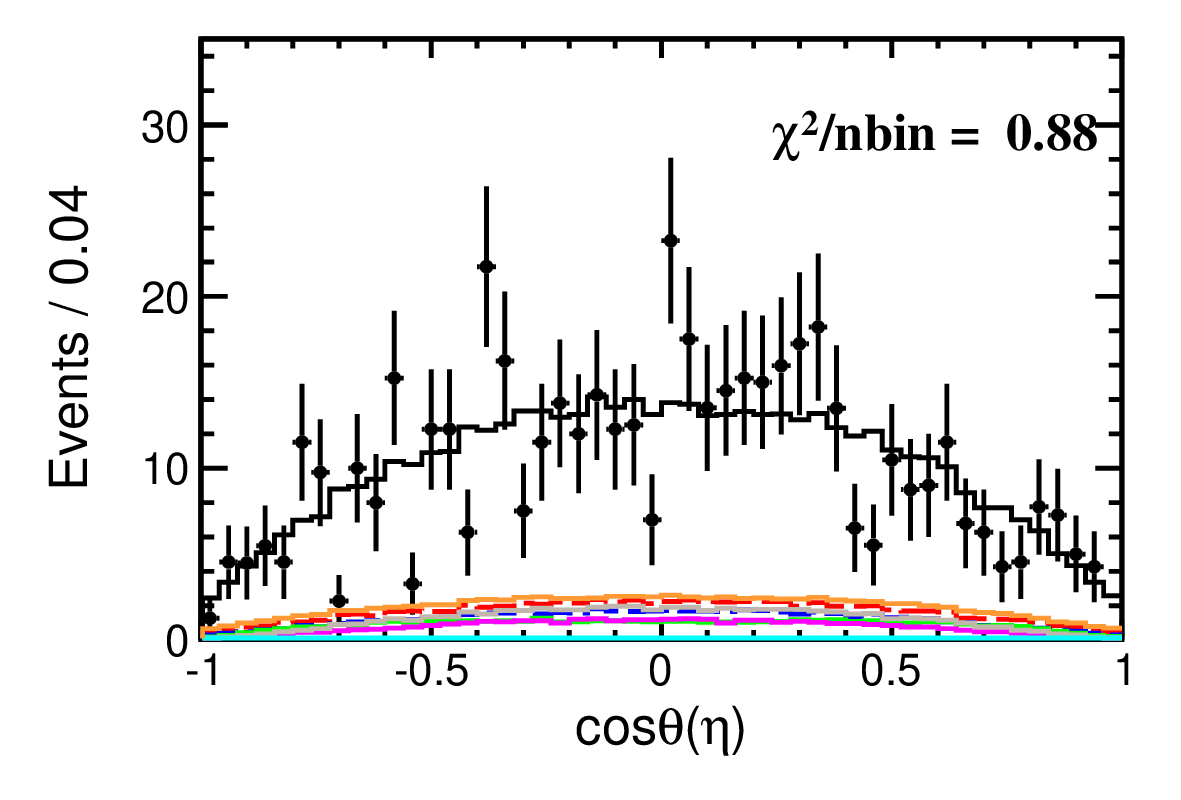}
 \put(80,48){ (f)}
 \end{overpic}
\setlength{\abovecaptionskip}{1pt}
 \caption{ Background-subtracted data (black dots) and PWA fit projections (black lines) for the invariant mass distributions of (a) $\phi\eta$, (b) $\phi\eta'$, and (c) $\eta\eta'$, and for the angular distributions of (d) cos$\theta$ of the $\eta$ in the $\phi \eta$ helicity frame, (e) cos$\theta$ of the $\eta'$ in the $\phi \eta'$ helicity frame, and (f) cos$\theta$ of the $\eta$ in the center-of-mass rest frame.}
 \label{PWAfiguure2125:2}
\end{figure*}

Figures~\ref{Fig:1}(d) shows the Dalitz plot for these events in the  signal regions.
Figure.~\ref{PWAfiguure2125:2} shows the projections of $\phi\eta$, $\phi\etap$, and $\eta\etap$ invariant mass ($M_{\phi\eta}$, $M_{\phi\etap}$ and $M_{\eta\etap}$) and the corresponding angular distributions, after background subtraction, for the candidate events.
%The background-subtracted invariant-mass distributions of the $\phi\eta$, $\phi\etap$, and $\eta\etap$ combinations ($M_{\phi\eta}$, $M_{\phi\etap}$ and $M_{\eta\etap}$) for the candidate events, as well as the corresponding angular distributions, are shown in Fig.~\ref{PWAfiguure2125:2}. 
A structure with a mass around 2.3~$\gevcc$ in the $M_{\phi\eta}$ and $M_{\phi\etap}$ distributions is clearly observed. To determine the properties of the newly observed structure, a PWA based on the GPUPWA framework~\cite{Berger:2010zza} is performed. 
Quasi two-body decay amplitudes in the three sequential decay processes $\psi(3686) \to \phi$X with X $\to \eta \eta'$; $\psi(3686) \to \eta$X with X $\to \phi \eta'$; and $\psi(3686) \to \eta'$X with X $\to \phi \eta$ are constructed using the covariant tensor amplitudes described in Ref.~\cite{Zou:2002ar}. The intermediate states are parameterized with a relativistic Breit-Wigner (BW) function with mass-dependent width. Parameters of the known intermediate states are fixed to the PDG values. The complex coefficients of the amplitudes (relative magnitudes and phases) and resonance parameters (mass and width) are determined by an unbinned maximum likelihood fit.
The probability of observing $N$ events  in the data sample is
 \begin{equation}
      S(\xi)= -\ln\mathcal{L} =
      -\sum_{i=1}^{N}\ln(\frac{|M(\xi_i)|^2\varepsilon(\xi_{i})}{\sigma'}),
 \label{PWAf:6}
 \end{equation}
 where $\xi_{i}$ stands for all kinematic variables in the $i$th event, $\varepsilon(\xi_{i})$ is the detection efficiency, and $M(\xi_i) = \sum A(\xi_i)$ is the matrix element describing the decay processes from $\psip$ to the final state $\phi\eta\etap$. In addition, $A(\xi_i)$ is the amplitude of the corresponding intermediate resonance and $\sigma'=\int d\Phi|M(\xi)|^2\varepsilon(\xi)$ is the normalization integral. 
The background is taken into account by subtracting normalized sideband and continuum events
from the total log likelihood value. The free parameters are optimized using MINUIT~\cite{James:1975dr}. 

In the fit procedure, all intermediate states reported by the PDG and which conserve quantum numbers $J^{PC}$ are considered.
Coherent non-resonant components are also taken into account by including very broad intermediate states in the fit.
The fitting procedure begins with a baseline solution that includes phase space~(PHSP) and the well-established $\phi(2170)$ and $h_{1}(1900$) states, and then adds all other possible intermediate resonances one by one.
The statistical significance of each resonance is evaluated by examining the change in the log likelihood value and the number of free parameters in the fit with and without the resonance included.
All components with a statistical significance greater than 5$\sigma$ are added to the baseline solution. 
After that, all the removed components are tested again to make sure that they still do not satisfy the 5$\sigma$ requirement.  It is found that this set of components does not adequately describe the structure around 2.3\,$\gevcc$ in the $\phi \eta$ and $\phi\etap$ invariant mass spectra, so an additional resonance is added to the fit.
To investigate the $J^{PC}$ of the additional resonance, different $J^{PC}$ combinations are tested and the $1^{+-}$ combination is found to produce
the largest change of the log likelihood value.
The final set of amplitudes therefore contains a significant contribution from an axial-vector state with $J^{PC}$ = $1^{+-}$, denoted as $X$(2300). Its significance is 9.6$\sigma$ and the corresponding mass and width are determined to be 2316 $\pm$ 9 $\mevcc$ and 89 $\pm$ 15 $\mev$ by a scan of the log likelihood value.

The PWA results with the baseline set of amplitudes, including the masses and widths of the resonances, the product branching fraction of each component and the statistical significances , are summarized in Table~\ref{summa:Br}. 

\renewcommand{\arraystretch}{1.3}
     \begin{table}[htbp]
    \caption{The obtained masses, widths, product branching fractions and statistical significances (Sig.)  for the $\psip$ decay and the subsequent intermediate state decay, where the first uncertainties are statistical, and the second systematic. The term ``N/A" denotes unavailability.}\label{summa:Br}
    \begin{center}
    \scriptsize
    \begin{tabular}{ l |c|c|c |c}
    \hline
    \hline
     Process &       M (MeV/$c^{2}$) &  $\Gamma$ (MeV)   & $\mathcal{B}$ ($10^{-6}$)  & Sig. \\ \hline
     $h_{1}(1900)\eta$~\cite{BESIII:MEMO}      &1911 $\pm$ 6 $\pm$ 14      &149 $\pm$ 12 $\pm$ 23   &  3.8 $\pm$ 1.4 $\pm$ 1.5 & 5.3$\sigma$  \\
     $\phi(2170)\eta$~\cite{ParticleDataGroup:2020ssz}      & \multirow{2}{*}{$2162 \pm 7 $} & \multirow{2}{*}{$100 ^{+31}_{-23} $} &  3.3 $\pm$ 1.4 $\pm$  1.2  & 5.1$\sigma$ \\
     $\phi(2170)\eta'$~\cite{ParticleDataGroup:2020ssz}  &        &                   &  3.8 $\pm$ 0.9 $\pm$     0.5  & 6.0$\sigma$ \\
     $X(2300)\eta$    &\multirow{2}{*}{$2316 \pm 9 \pm 30$} &\multirow{2}{*}{$89 \pm 15 \pm 26$ }  &  2.2 $\pm$ 0.7 $\pm$ 0.7 & 5.6$\sigma$  \\

     $X(2300)\eta'$   & &   & 4.8 $\pm$ 1.3  $\pm$  0.7 &9.6$\sigma$  \\
     PHSP & N/A & N/A &  6.1 $\pm$ 2.3  $\pm$  1.8 & 5.2$\sigma$ \\
    \hline \hline
    \end{tabular}
    \end{center}
    \end{table}

Figure~\ref{PWAfiguure2125:2} shows the invariant mass distributions of $\phi\eta$, $\phi\eta'$, and $\eta\eta'$ for the data, the PWA fit projections and the angular distribution of $\cos\theta(\eta)$, $\cos\theta(\eta')$, and $\cos\theta(\phi)$ in the corresponding helicity frame. 

The signal efficiency of $\psi(3686) \to \phi \eta \eta' $ is determined to be 9.88\% using the MC sample weighted according to the PWA result. The total branching fraction of $\psi(3686) \to \phi \eta \eta' $ is calculated to be (3.14 $\pm$ $0.17_{\mathrm{stat}}$ $\pm$ $0.24_{\mathrm{syst}}$)$\times$ 10$^{-5}$, where the first uncertainty is statistical and the second is systematic.

Sources of systematic uncertainties are considered in two categories, non-PWA-related and PWA-related systematics. The first category includes the uncertainties of photon detection efficiency (1.0\% per photon)~\cite{BESIII:2010ank}, MDC tracking efficiency (1.0\% per charged track)~\cite{BESIII:2011wmh},  PID efficiency (1.0\% per charged track)~\cite{BESIII:2011ysp}, 4C kinematic fit (1.3\%),  $\phi$, $\eta$ and $\eta'$ mass windows (1.5\%, 0.2\%, 0.6\%), quoted branching fractions (1.8\%)~\cite{ParticleDataGroup:2020ssz}, total number of $\psi(3686)$ events (0.5\%), MC statistics (0.3\%) and continuum background (1.6\%). 
 The uncertainty related to the 4C kinematic fit is estimated by correcting the helix parameters of the simulated charged tracks to match the resolution in the data~\cite{BESIII:2012mpj}. The difference between the signal efficiencies with and without correction is regarded as the systematic uncertainty. The systematic uncertainties from the mass resolution are estimated by fitting the mass spectrum with the MC shape convolved with a fixed-parameter Gaussian function, where the fixed parameters are extracted from control samples of $\psip \to \phi\pipi$ and $\psip \to \omega\eta\etap$. The resulting changes on the signal yields are taken as the systematic uncertainty.
The second category, which includes the uncertainties from the resonance description, resonance parameters, extra resonances, the Blatt Weisskopf barrier factor~\cite{barriersys1,barriersys2} whcich is included in the PWA decay amplitudes, and background uncertainty, which all arise from the PWA fit procedure, affects the measurements of both the branching fraction and resonance parameters. To estimate these uncertainties, alternative fits with different scenarios are performed, and the resulting changes on the branching fractions and the resonance parameters are taken as the systematic errors.
Uncertainties from the BW parameterization are estimated by replacing the mass-dependent-width BW with the constant width. 
Uncertainties from the resonance parameters are estimated by fitting with masses and widths fixed to the values randomly generated from Gaussian distributions with their uncertainties taken from the PDG and the standard deviation of the results is taken as the uncertainty. 
The uncertainties related to extra resonances are estimated by adding known resonances with significances greater than 3$\sigma$  and an additional possible structure around 2.6 $\gevcc$ with $J^{PC} = 1^{+-}$.
The largest change from the nominal PWA fit is taken as the uncertainty.  
Uncertainties due to the
barrier factor~\cite{barriersys1,barriersys2} are estimated by varying the radius of the centrifugal barrier from 0.7 to 1.0 fm.
The uncertainty associated with the background description is estimated by using different sideband regions, and changing the background level by varying the sideband normalization factors by one standard deviation.

  In summary, using (2712.4~$\pm$~14.3)$~\times$~10$^{6}$  $\psi(3686)$ events collected by the BESIII detector, a PWA of the decay $\psi(3686) \to \phi \eta \etap$ is performed. The analysis results in the first observation of an axial-vector state around 2.3\,$\gevcc$. Its mass and width are measured to be $2316\pm9_{\mathrm{stat}}\pm30_{\mathrm{syst}}\, \mevcc$ and $89 \pm 15_{\mathrm{stat}} \pm26_{\mathrm{syst}}\,\mev$, respectively. The product branching fractions of $\mathcal{B}(\psi(3686) \to X(2300) \eta') \mathcal{B}(X(2300)\to \phi \eta)$ and $\mathcal{B}(\psi(3686) \to X(2300) \eta)\mathcal{B}(X(2300)\to \phi \eta')$ are determined to be (4.8~$\pm~1.3_{\mathrm{stat}}~\pm~0.7_{\mathrm{syst}})~\times ~10^{-6}$ and (2.2~$\pm~0.7_{\mathrm{stat}}~\pm~0.7_{\mathrm{syst}})~\times~10^{-6}$, respectively, where the first uncertainty is statistical and the second is systematic. The branching fraction $\mathcal{B}(\psi(3686) \to \phi \eta \eta')$ is measured for the first time to be (3.14~$\pm~0.17_{\mathrm{stat}}~\pm~0.24_{\mathrm{syst}})\times10^{-5}$.

{%\color{red}
%In the spectrum of axial-vector strangeonium states, only the ground axial-vector state $h_1(1380)$~\cite{CrystalBarrel:1997kda,BESIII:2015vfb,BESIII:2018ede} has been confrimed in PDG~\cite{ParticleDataGroup:2020ssz}, and excited states, e.g., $h_1(2P)$ and $h_1(3P)$, have been predicted by theory~\cite{Barnes:2002mu,Li:2020xzs,h13p_1,h13p_2,h13p_3,h13p_4} but have not been observed by experiment. Recently, BESIII observed the $1^{+-}$ state, which is a candidate for the $h_1(2P)$ state~\cite{BESIII:MEMO}. However, the $h_1(3P)$ still remains unobserved by experiments. Decay modes such as $\phi\eta$ and $\phi\eta'$ are proposed as the dominant channels to search  $h_1(3P)$. 

Theoretical predictions regarding parameters of the axial-vector $s\bar{s}$ state $h_{1}$(3P) and the 1S wave multiplets $T_{(ss\bar{s}\bar{s})1^{+-}}$ are shown in Table~\ref{summa:Br1}.
%Table~\ref{summa:Br1} shows the theoretical predictions about the 1S wave multiplets $T_{(ss\bar{s}\bar{s})1^{+-}}$ and the axial-vector $s\bar{s}$ state $h_{1}$(3P).
}
{%\color{red} 
Comparing to the theoretical predictions about $h_{1}$(3P), 
the mass of the observed resonance deviates from that of $h_{1}(3P)$.
There is a noticeable discrepancy in the mass of $T_{(ss\bar{s}\bar{s})1^{+-}}$ from the theoretical expectations, making it challenging to determine the consistency with our results.
%For $T_{(ss\bar{s}\bar{s})1^{+-}}$, discrepancy on the mass of the above theoretical expectations is observed,  hard to judge the consistent with our results.
%None of the above theoretical expectations are in good agreement with our experimental results.
The $X(2300)$ observed in this work provides critical information for understanding the axial-vector strangeonium spectrum and $T_{ss\bar{s}\bar{s}}$.}

     \begin{table}[htbp]
    \caption{Comparison of experimental measurements and theoretical predictions for the mass and width of the $h_{1}(3P)$ and $T_{ss\bar{s}\bar{s}}$ resonances. The term ``N/A" denotes unavailability.}\label{summa:Br1}
    \begin{center}
    \scriptsize
    \begin{tabular}{ l |c|c }
    \hline
    \hline
     Resonance &       M (MeV/$c^{2}$) &  $\Gamma$ (MeV)   \\ \hline
     $h_{1}(3P)$     &2435~\cite{Li:2020xzs}     &269 ~\cite{Li:2020xzs}   \\
     $h_{1}(3P) $     & 2449~\cite{h13p_1}& N/A \\
     $h_{1}(3P) $     & 2100~\cite{Chen:2015iqa,Wang:2019qyy}& N/A \\
     $h_{1}(3P) $  &  2490~\cite{h13p_2}   &  N/A   \\
     $h_{1}(3P) $   &2398~\cite{h13p_3} & N/A  \\

     $h_{1}(3P) $  &2495.51$\pm$1.46~\cite{h13p_4} & N/A    \\
     $h_{1}(4P) $  & 2340~\cite{Chen:2015iqa,Wang:2019qyy}& N/A \\
     \hline
     $T_{(ss\bar{s}\bar{s})1^{+-}}$  & 2323~\cite{tetraquark:2021} & N/A  \\
     $T_{(ss\bar{s}\bar{s})1^{+-}}$  & 1960~\cite{Lu:2019ira} & N/A  \\
     $T_{(ss\bar{s}\bar{s})1^{+-}}$  & $2000^{+100}_{-90}$~\cite{Cui:2019roq} & N/A  \\
     \hline
     This work   & 2316  $\pm$ 9 $\pm$ 30 & 89 $\pm$ 15 $\pm$ 26  \\
    \hline \hline
    \end{tabular}
    \end{center}
    \end{table}

%%%%%%%%%Summary%%%%%%%%%%%%%%%%%%%%%%%%%%%%%%%%%%%%%%%%%%%%%%%%%%%%%%%%%%%%%%%%%%%%%%%%%%%%%%%%%%%%%%%%%%%%%%%%%%%%%%%%%%
The authors would like to thank Prof.~Xian-Hui Zhong and Prof.~Xiang Liu for their valuable theoretical discussions.
The BESIII Collaboration thanks the staff of BEPCII, the IHEP computing center and the supercomputing center of USTC for their strong support. This work is supported in part by National Key R\&D Program of China under Contracts Nos. 2023YFA1609400, 2020YFA0406400, 2020YFA0406300; National Natural Science Foundation of China (NSFC) under Contracts Nos. 11635010, 11735014, 11835012, 11935015, 11935016, 11935018, 11961141012, 12022510, 12025502, 12035009, 12035013, 12192260, 12192261, 12192262, 12192263, 12192264, 12192265, 11335008, 11625523, 12035013, 11705192, 11950410506, 12061131003, 12105276, 12122509, 12225509, 12205255; The Chinese Academy of Sciences (CAS) Large-Scale Scientific Facility Program; Joint Large-Scale Scientific Facility Funds of the NSFC and CAS under Contract No. U1832207, U1732263, U1832103, U2032111, U2032105; The CAS Center for Excellence in Particle Physics (CCEPP); 100 Talents Program of CAS; The Institute of Nuclear and Particle Physics (INPAC) and Shanghai Key Laboratory for Particle Physics and Cosmology; ERC under Contract No. 758462; European Union's Horizon 2020 research and innovation programme under Marie Sklodowska-Curie grant agreement under Contract No. 894790; German Research Foundation DFG under Contracts Nos. 443159800, Collaborative Research Center CRC 1044, GRK 2149; Istituto Nazionale di Fisica Nucleare, Italy; Ministry of Development of Turkey under Contract No. DPT2006K-120470; National Science and Technology fund; National Science Research and Innovation Fund (NSRF) via the Program Management Unit for Human Resources \& Institutional Development, Research and Innovation under Contract No. B16F640076; Olle Engkvist Foundation under Contract No. 200-0605; STFC (United Kingdom); Suranaree University of Technology (SUT), Thailand Science Research and Innovation (TSRI), and National Science Research and Innovation Fund (NSRF) under Contract No. 160355; The Royal Society, UK under Contracts Nos. DH140054, DH160214; The Swedish Research Council; U. S. Department of Energy under Contract No. DE-FG02-05ER41374.

%%%%%%%%%%%%%%%%%%%%%%%%%%%%%%%%%%%%%%%%%%%%%%%%%%%%%%%%%%%%%%%%%%%%%%%%%%%%%%

\end{document}